13.00.00 Pedagogical science

13.00.00 Педагогические науки

# The Effects of Computer-assisted and Distance Learning of Geometric Modeling

[1] Omer Faruk Sozcu
[2] Rushan Ziatdinov
[3] Ismail Ipek

[1-3] Department of Computer & Instructional Technologies,
Fatih University, 34500 Büyükçekmece, Istanbul, Turkey
[1] PhD in Education, Assistant Professor
E-mail: omersozcu75@gmail.com
[2] PhD in Mathematical Modeling, Assistant Professor
E-mail: rushanziatdinov@gmail.com, ziatdinov@fatih.edu.tr
URL: www.ziatdinov-lab.ru
[3] PhD in Education, Associate Professor
E-mail: ismailipek34@gmail.com

**ABSTRACT.** The effects of computer-assisted and distance learning of geometric modeling and computer aided geometric design are studied. It was shown that computer algebra systems and dynamic geometric environments can be considered as excellent tools for teaching mathematical concepts of mentioned areas, and distance education technologies would be indispensable for consolidation of successfully passed topics.

## INTRODUCTION

Information Technology has created revolutionary changes in education. The advantages of e-learning are twofold: we can overcome the restrictions of time or space and we can study individually and cooperatively. In e-learning, collaborative learning is the main learning method, whereby various students study by using various learning materials in a common learning space to gain the effects of synergy (Alavi, 1994).

One of the most important tasks in mathematics and geometry education today is the revision of curricula and teaching methods to take advantage of electronic information technology. Developments in this decade alone have presented us with inexpensive and powerful hardware and software tools that challenge every traditional assumption about what we should teach, how we should teach and what students can learn. Today, one of the pieces of instructional software to teach math contents is GeoGebra, which can be used in distance education as an e-learning tool. The program is made up of interactive geometry software that also offers algebraic possibilities, such as entering equations directly. It is aimed at students (aged 10 to 18) and teachers in secondary schools (Hohenwarter & Fuchs, 2004).

The World Wide Web provides great opportunities for creating virtual classrooms for learners and instructors involved in distance education. Many software environments take advantage of client-server communication on the Internet and support open and distance learning. Distance learning has greatly improved in the past few years, as both students and educators have become more comfortable with the technology and as stories of the best practices have been shared and duplicated.

Technical subjects, such as mathematics, do not transfer easily to the distance education mode, so teaching and learning in such environments present special challenges. Any technologies employed for this purpose must satisfy the need for visual material, algebraic symbolism and geometric representation. There should be opportunities to explore ideas, to work on problem solving and to provide specific feedback, preferably through prompt verbal and visual explanations. Most importantly, the medium should also facilitate the development of motivation and enthusiasm for the subject and the potential for interactivity.

Geometric modeling is an interdisciplinary field of study, where applied mathematics and computer science intersect. It studies specific methods and algorithms for the mathematical description of curves and surfaces and their applications in computer-aided design (CAD), shape modeling and industrial geometry. Nowadays most geometric modeling is done with computers and is used in computer-based applications. Two-dimensional models are important in computer typography, technical drawing and mobile robot path planning. Three-dimensional models are central to computer-aided design and manufacturing (CAD/CAM) and are widely used in many applied technical fields, such as civil and mechanical engineering, architecture, geology, computer vision and medical image processing. Since geometric modeling is highly connected with scientific visualizations and animations, there is a need for specific methodologies of computer assisted learning and distance education. Computer algebra systems (CAS) and dynamic geometry environments (DGS) are well-known as the most effective tools for teaching geometric modeling. They provide plenty of instruments for scientific visualization and modeling and in conjunction with distance education tools can be considered as a powerful technique for teaching.

The World Wide Web provides great opportunities for creating virtual classrooms of learners and instructors involved in distance education. Many software environments take advantage of the client-server communication on the Internet and support open and distance learning. Educational research shows that monitoring the students' learning is an essential component of high quality education and is one of the major factors differentiating effective schools and teachers from ineffective ones. This also applies to online teaching, e.g., Helic et al. argue that good online tutoring requires the monitoring of a learner's progress with the material and testing of the acquired knowledge and skills on a regular basis. Assessment and measurement enable teachers to gauge the student response, feedback and progress towards goals and is crucial in distance education where the learner is impaired by the lack of casual contact with the teacher and other students. Technical subjects such as mathematics do not transfer easily to distance education mode, so teaching and learning in such environments present special challenges. Any technologies employed for this purpose must satisfy the need for visual material, algebraic symbolism and geometric representation. There should be opportunities to explore ideas, to work on problem solving and to provide specific feedback, preferably through prompt verbal and visual explanations. Most importantly, the medium should also facilitate the development of motivation, enthusiasm for the subject and the potential for interactivity. An interactive teaching and learning model involving Desktop Video Conferencing (DVC) and other audio-graphic facilities were developed and trialed for distance education in undergraduate geometry and mathematics. It appears that very little has been reported previously in this area of mathematics teaching, certainly not on the scale of this development. There is a possibility of integrating the virtual learning environments (VLE) with computer applications and graphics software to enable the difficult task of representing algebraic, geometric and numeric concepts, all of which are essential for the development of higher level mathematical topics in distance learning.

At this time, different learning environments can be used as components of learning techniques in distance learning by using geometric modeling with GeoGebra. The software system provides benefits in problem or project based learning, case-based learning and information sources for solving problems, cognitive construction tools, learning with collaboration and social or contextual support (Jonassen, Peck & Wilson, 1998). These learning environments can be effective for learning math and geometry in distance education as well. As a result, geometric modeling has vital effects on the learning process, which can be a part of computer-based and distance learning. These learning techniques are also very important variables in distance and computer assisted learning as types of dynamic and virtual learning.

**GeoGebra and its applications in mathematics and geometric modeling**

GeoGebra is an interactive geometry, algebra and calculus application, intended for teachers and students. The possibilities of using DGS in university level mathematics education were studied in (Ziatdinov, 2009, 2010, 2012) and its pedagogical implications were discussed in (Karadag, 2011). Gulseçen et al. (2012) have investigated the views of students and their teacher's on the use of GeoGebra, and concluded that it is an excellent way to focus students' understanding in math topics. Ziatdinov & Rakuta (2012) have discussed a number of issues and problems associated with the use of computer models in the study of geometry in university, as well as school mathematics, in order to improve its efficiency. They also pointed out that computer models for different geometrical problems should be widely used for increasing the efficiency of teaching and learning. Francisco Pérez-Arribas (2012) has studied the possibilities of instructing naval architects with GeoGebra and his methodology has been in use since 2007.

In this research we are interested in applying GeoGebra for a geometric modeling course and seeing its powerful, as well as weak, features. The advantages of GeoGebra can be stated as follows.
• Simple graphical interface;
• Simplicity in inputting equations of parametric curves, such as Bézier, B-splines, or NURBS and other polynomial curves (Farin, 2002);
• The possibility of using a shape parameter and modifying the shape of the curve using a slider;
• Modifying the shape of the curve by changing the positions of its control points.

On the other hand there are many problems to be solved, for which GeoGebra is useless. Its disadvantages may be summarized as:
• Impossible to use for computing non-linear curve segments such as, log-aesthetic curves (Miura, 2006; Yoshida & Saito, 2006; Ziatdinov et al., 2012), multispirals (Ziatdinov et al., 2012) and superspirals (Ziatdinov, 2012) because of a curve's complicated representations in terms of special functions;
• Impossible to deal with parametric surfaces, which can be done in Rhinoceros 3D, which is commercial NURBS-based 3-D modeling software;
• Some other problems, which are probably connected with program bugs (Fig. 1).

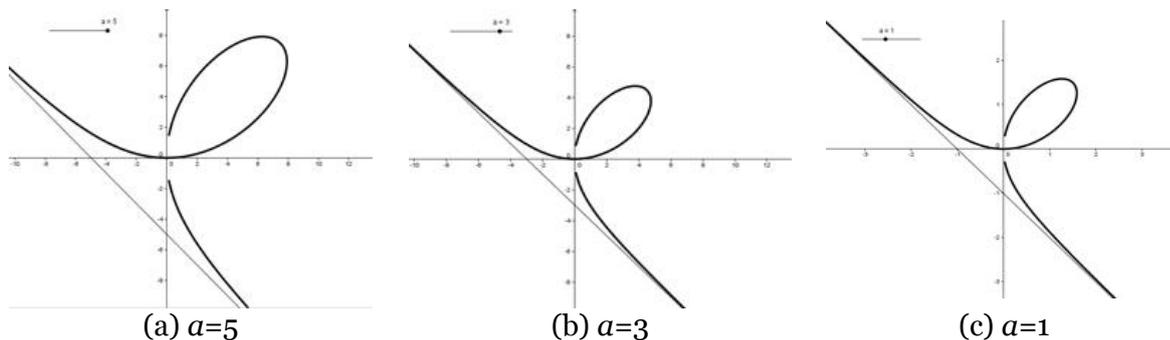

(a) $a$=5         (b) $a$=3         (c) $a$=1

Figure 1. The Folium of Descartes ($x^3 + y^3 - 3axy = 0$) with different values of parameter $a$. As it can be seen from the pictures, a curve segment at the neighborhood of the origin is not drawn

## Geometric modeling with Computer Algebra Systems

A computer algebra system (CAS) is a software program that facilitates symbolic mathematics. The core functionality of a CAS is the manipulation of mathematical expressions in symbolic form. CAS is a very powerful tool for teaching and research in geometric modeling problems; as well as being traditional symbolic manipulations, they are programming languages, allowing users to implement their own algorithms. CAS also includes arbitrary-precision numeric operations, APIs for linking it to an external program such as a database, or using in a programming language to use the computer algebra system, add-ons for use in applied mathematics such as physics, bioinformatics, computational chemistry and packages for physical computation, graphic production and editing such as computer generated imagery and sound synthesis.

CAS can also be considered as an excellent tool for computational science (or scientific computing), which is concerned with constructing mathematical models and quantitative analysis techniques and using computers to analyze and solve scientific problems. That is why we think that CAS (Maple, Mathematica, Matlab, Derive) is one of the best tools for teaching geometric modeling. Below some Maple examples on computing free-form curves are shown.

*Example 1. Bézier curve*

A Bézier curve is a parametric curve frequently used in computer graphics and related fields. Its Maple code is presented below.

```
> bezier_curve:=proc(points)
  local n,m,P_x,P_y,B,x,y,p0,p1,p2,k,graph1,graph2:
  with(plots):
  with(plottools):
  n:=nops(points)-1;
  for m from 1 to n+1 do
  points[m]
  end do;
  P_x:=(i)- points[i+1][1];
  P_y:=(i)- points[i+1][2];
  P_x(1);
  P_y(1);
  B:=(i,t)- binomial(n,i)*(1-t)^(n-i)*t^i;
  x:=(t)-  sum(P_x(i)*B(i,t),i=0..n);
  print(X(t)=x(t));
  x(t);
  y:=(t)- sum(P_y(i)*B(i,t),i=0..n);
  print(Y(t)=y(t));
  y(t);
  p0:=plottools[curve](points, color=red, linestyle=dash, thickness=2):
  p1:=plots[pointplot](points,symbol=solidcircle,symbolsize=20,color=blue):
  p2:=plot([x(t),y(t),t=0..1],color=black,      thickness=4,      font=[TIMES,      ROMAN,
  30],title="Bezier          curve",titlefont=[TIMES,  ROMAN,  20],axes=boxed,labels=["x(t)",  "y
  (t)"],resolution=400,scaling=constrained);
  graph1:=plots[display]({p0,p1,p2});
  print(graph1);
  k:=(t)-                                              (diff(x(t),t)*diff(y(t),t$2)-
  diff(y(t),t)*diff(x(t),t$2))/(diff(x(t),t)^2+diff(y(t),t)^2)^(3/2);
  graph2:=plot(k(t),t=0..1,color=black,       thickness=4,       font=[TIMES,       ROMAN,
  30],title="Curvature      plot",titlefont=[TIMES,    ROMAN,    20],axes=boxed,labels=["t",    "k
  (t)"],resolution=400,scaling=constrained);
  print(graph2);
  end proc:
> LibraryTools[Save](bezier_curve, "geom_modelling.m"):
```

This example created by us was saved in the "geom_modeling.m" file. By inputting the curve's control points as is shown below, the curve's graph and its curvature plot can be created (Fig. 2).

```
> restart;
> points := [[0,0],[0.5,2],[2,5],[5,3],[6,0]];
> bezier_curve(points);
```

For more comprehensive information on using Maple and Matlab for modeling curves and surfaces the reader is referred to the books of Vladimir Rovenski (2000, 2010).

**Math and Geometry learning in distance learning and CBL**

In the math and geometry learning process, game learning can be designed as a type of e-learning in distance education and computer based-learning. When we talk about distance education, integrated e-learning tools can be defined as e-mail, CDs, interactive video and other new technologies for learning. Geometric modeling and its software tools, which include several functions for learning geometry, indicate good teaching strategies by linking computers. The process also uses computers for effective learning and teaching. For this reason, geometric modeling creates instructional opportunities for learners in CBL and distance learning.

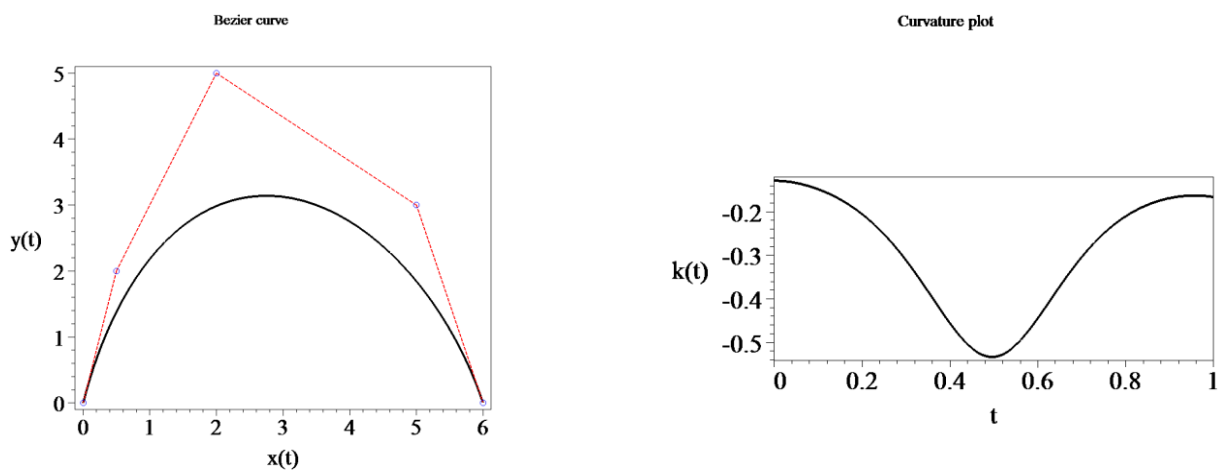

Figure 2. The plot of Bézier curve and its curvature function

Instructional designers and technologists should select this technology to reach their objectives and achievements in schools. At this time, distance learning with geometric modeling provides techniques for students and teachers in math by using new technologies and e-learning tools. Distance learning has advantages in learning environments, as well as traditional instructions. They can be defined as follows.

Geometric modeling has easy accessibility whenever a user needs some information from the trainer center. The model also has no waste time for each class in distance learning as synchronous or asynchronous learning style. A teacher can use the GeoGebra system or software whenever he wants and study in any convenient location with an internet connection. Self-paced learning includes quickly browsing materials and studying materials at a personal speed and intensity with GeoGebra. In addition, it provides flexibility to join conversations in the bulletin board discussion areas at any hour and to review your classmates' comments since the previous visit. All activities in distant learning or CB can be produced by use of GeoGebra system and its contents, in order to teach math concepts for students.

**CONCLUSION**
In this paper we have presented the effects of Geometric modeling, such as the GeoGebra system or software, in distance learning. CBL Geometric modeling can also work as an e-learning tool for students because it teaches several concepts in math such as line, animation, graphics and intersections in geometry, as well as problem solving strategies. In addition, e-learning teaching strategies have worked effectively in the learning process when teaching contents described in the GeoGebra system and at different grades. The software provides links between contents and students' activities. Using hypertext and hypermedia techniques in distance learning and CBL has also been considered. A further idea is that there is a node (s) and link (s), which represent interactivity and its effects on learning. This learning environment gives us opportunities for using technologies as part of distance learning and computers. Mean and geometric modeling by algebra or GeoGebra learning systems can be effectively used for learners at different ages. For this reason, there is a simple graphical interface, parametric curves, points and other polynomial curves with 3D modeling software, which are connected with program bugs. These learning activities by geometric modeling can develop learners' skills and achievements. The learning process also includes visualization, communications, visual learning, games, animations and high level cognitive learning skills for learning in distance education. As a result, distance learning software and CBL techniques can be used to teach math and geometry concepts by geometric modeling, as well as conventional CBL or Computer-based Instruction (CBI). Thus, Geometric modeling software or the GeoGebra software can be used effectively for future classes instead of CBI programs in distance learning to teach programming with computer algebra systems.

УДК 530

**Эффекты компьютерного и дистанционного обучения геометрическому моделированию**

[1] Омер Фарук Созджу
[2] Рушан Зиатдинов
[3] Исмаил Ипек

[1-3] Университет Фатих, Турция
34500 Буюкчекмедже, Стамбул
[1] доктор педагогических наук, Ассистент-Профессор
E-mail: omersozcu75@gmail.com
[2] кандидат физико-математических наук, Ассистент-Профессор
E-mail: rushanziatdinov@gmail.com, ziatdinov@fatih.edu.tr
URL: www.ziatdinov-lab.ru
[3] доктор педагогических наук, Ассоциированный Профессор
E-mail: ismailipek34@gmail.com

**Аннотация.** В работе рассматриваются эффекты компьютерного и дистанционного обучения геометрическому моделированию и компьютерному геометрическому дизайну. Также демонстрируется, что системы компьютерной математики и интерактивные геометрические среды являются замечательными средствами для обучения математическим основам упомянутых дисциплин, а технологии дистанционного обучения незаменимы при закреплении пройденного материала.

**Ключевые слова:** виртуальное обучение; интерактивное обучение; геометрическое моделирование; визуализации; автоматизированное геометрическое проектирование; система компьютерной математики; интерактивная геометрическая среда; GeoGebra; Maple; CAD; кривая Безье; дистанционное образование.